\documentclass[conference]{IEEEtran}
\IEEEoverridecommandlockouts

\usepackage{cite}
\usepackage{epstopdf}
\usepackage{amsmath,amssymb,amsfonts}
\usepackage{algorithmic}
\usepackage{graphicx}
\usepackage{textcomp}
\usepackage{xcolor}
\usepackage{subcaption}
\usepackage{booktabs}
\usepackage[linesnumbered,ruled]{algorithm2e}

\def\BibTeX{{\rm B\kern-.05em{\sc i\kern-.025em b}\kern-.08em
    T\kern-.1667em\lower.7ex\hbox{E}\kern-.125emX}}

\begin{document}
	\title{Beyond Dedicated-Active: A General Reliability Provisioning Framework for SFC Placement in Fog Computing}
	\author{
		\IEEEauthorblockN{
			Negin Doostar\textsuperscript{1},
			Mohammad Reza Heidarpour\textsuperscript{1},
			and Amir Khorsandi\textsuperscript{1}
		}
		\IEEEauthorblockA{
			\textsuperscript{1}Department of Electrical and Computer Engineering\\
			Isfahan University of Technology\\
			Isfahan, Iran\\
		}
	}
	\maketitle
\begin{abstract}
The explosive growth of Internet of Things (IoT) devices has strained traditional cloud infrastructures, highlighting the need for low-latency and energy-efficient alternatives. Fog computing addresses this by placing computation near the network edge. However, limited and heterogeneous fog resources pose reliability challenges, especially for mission-critical applications.
On the other hand, to improve flexibility, applications are deployed as Service Function Chains (SFCs), where each function runs as a Virtual Network Function (VNF). While scalable, this approach is more failure-prone than monolithic deployments, necessitating intelligent redundancy and placement strategies.
This paper addresses the reliability-aware SFC placement problem over heterogeneous fog servers through the lens of reliability theory. We explore four redundancy strategies, combining shared vs. dedicated and active vs. standby modes, and propose a general framework to minimize latency and cost while meeting reliability and deadline constraints. The problem is formulated as an Integer Non-Linear Program (INLP), and two genetic algorithm (GA)-based solutions are developed.
Simulation results show that shared-standby redundancy outperforms the conventional dedicated-active approach by up to 84\%. 
\end{abstract}
\section{INTRODUCTION}
The increasing proliferation of Internet of Things (IoT) devices generates massive datasets, straining the capabilities of traditional cloud computing architectures.  Fog computing has emerged as a promising solution, strategically positioning computational resources and services closer to the network edge, between end-users and centralized clouds. This decentralized paradigm mitigates the limitations of transmitting large IoT datasets to distant cloud infrastructures, specifically addressing bandwidth constraints and high energy consumption. Moreover, by processing and storing data at the network edge, fog computing enables faster response times and enhanced proximity services compared to traditional cloud models.
However, fog computing is not a panacea and comes with its own set of challenges, including resource limitations and heterogeneity, among others \cite{r1}.

To address the dynamic and diverse requirements of IoT applications in cloud-fog environments, an effective solution is to deploy services in the form of ``Service Function Chains'' (SFCs) \cite{r2}.
Unlike the monolithic approach, in the SFC method,  the application is  decomposed into a series of interconnected components. Each component can be deployed as a ``Virtual Network Function'' (VNF) on cloud or fog servers. This approach admits scalability and placement flexibility  for individual components \cite{r3}.
On the other hand, the downside of SFC over cloud and specially fog servers is higher delay and  failure rate compared to the monolithic method where the whole software is tightly coupled with a dedicated hardware. As a result, for industrial and mission critical applications such as health care and autonomous vehicles with  ultra-reliable (e.g., 99.999\% availability) and low latency demands, SFC implementation must be safeguarded through  redundancy provisioning and deliberate VNF placement.
Finding the optimal balance between redundancy reservation on one side and the challenges of resource scarcity and power limitations on the other requires innovative approaches.
\\\indent
This paper seeks to examine the SFC reliability challenge through the lens of the well-established ``reliability theory'' aiming to introduce fresh perspectives and innovative strategies into the discussion. 
Reliability theory is a branch of probability with the focus on systems' failure analysis, using redundancy  to mitigate the failures and probabilistic modeling to predict system behavior \cite{r19}. 
Drawing on insights from reliability theory, this paper presents four distinct strategies, each offering different settings for the access mode (either dedicated or shared) and the operational state (either active or standby) of the backup nodes.
Accordingly, a general  framework is proposed to address the SFC placement problem across heterogeneous fog servers. This framework aims  to simultaneously minimize both latency and operational/maintenance costs while satisfying  reliability and deadline constraints.
Our main contributions are summarized as follows.
\begin{itemize}
	\item 
	This work advances redundancy provisioning strategies in SFC resource allocation. To our knowledge, prior studies have not explored redundancy sharing among VNFs or the use of standby mode for reserved resources. 
	\item 	The SFC placement problem across heterogeneous fog servers is formulated as an integer nonlinear programming (INLP), which permits various redundancy provisioning strategies for different SFCs while jointly optimizing average delay and deployment cost.
	\item 
	 We propose two genetic algorithm (GA)-based solutions to the problem of reliability-aware SFC placement. 
	\item 
	Numerical experiments compare various redundancy strategies and algorithms, showing performance improvements of up to 80\% over benchmark solutions. Moreover, results reveal that redundancy strategy significantly impacts performance and should be a key consideration. For instance, the shared-standby strategy outperforms the dedicated-active approach by 84\% in some scenarios.
\end{itemize}
The remainder of this article is structured as follows: Section II discusses related work. Section III demonstrates the importance of backup allocation strategies for ensuring reliability and then details various backup allocation strategies for SFCs. Section IV presents an INLP cost-aware formulation for latency-aware and reliable SFC placement. Section V discusses the proposed metaheuristic algorithms. Section VI evaluates and compares the performance of the proposed solutions under different strategies and benchmarking against existing approaches. Finally, Section VII concludes the article.

\section{RELATED WORKS}
\begin{table*}
	\begin{center}
		\caption{ Comparison of previous related works with the problem addressed in this study}
		\begin{tabular}{|c|*{4}{c|}c|c|c|c|}
			\hline
			& \multicolumn{4}{c|}{Reliability Strategy$ ^* $} & \multicolumn{4}{c|}{Optimization Criterion} \\\hline
			\cline{2-5}
			& DA & DS & SA & SS & Latency & Costs & Performance & Heterogeneous resources \\
			\hline
			
			\cite{r5} & \checkmark & & & & & & \checkmark & \\\hline
			\cite{r6} & \checkmark & & & & & \checkmark & & \\\hline
			\cite{r7} & \checkmark & & & & & \checkmark & \checkmark & \\\hline
			\cite{r8} & \checkmark & & & & \checkmark & \checkmark &  & \\\hline
			\cite{r9} & \checkmark & & & & & \checkmark & & \\\hline
			\cite{r10} & \checkmark & & & & \checkmark & \checkmark & & \checkmark \\\hline
			\cite{r14} & & & & & \checkmark & \checkmark & \checkmark & \checkmark \\\hline
			\cite{r15} & & & & & \checkmark & \checkmark &  & \checkmark \\\hline
			\cite{r16} & & & & & \checkmark  & \checkmark &  & \checkmark \\\hline 
			\cite{r17} & \checkmark & & & & & & \checkmark & \checkmark \\\hline
			\cite{r11} & \checkmark & & & & \checkmark & & & \checkmark \\\hline
			\cite{r13} & \checkmark & & & & \checkmark & \checkmark & & \\\hline
			\cite{r12} & \checkmark & & & & & \checkmark & &
			\\\hline
			This study & \checkmark & \checkmark & \checkmark & \checkmark & \checkmark & \checkmark & \checkmark & \checkmark \\\hline
			
		\end{tabular}
		\label{T1}
	\end{center}  
\begin{center}
	\parbox{0.7\textwidth}{\footnotesize $ ^* $  DA: Dedicated-Active, DS: Dedicated-Standby, SA: Shared-Active, and SS: Shared-Standby }
\end{center}
\end{table*}	
Enhancing network reliability necessitates addressing hardware and software failures, for which backup resource allocation is a key strategy. In SFCs, backups are often dedicated to individual VNFs, either actively running (dedicated-active) or on standby nodes (dedicated-standby). Alternatively, backup resources can be shared among multiple VNFs or the entire SFC, again either active (shared-active) or on standby nodes (shared-standby). The reliability of SFCs significantly impacts VNF placement decisions, affecting cost and quality of service (QoS). Numerous studies have explored methods to improve reliability through optimized SFC placement and backup resource allocation strategies, which will be elaborated upon.

In \cite{r5}, a two-stage approach is proposed for SFC placement considering backup allocation to ensure reliability. The first stage employs a heuristic to determine the minimum backup resources required without introducing significant delay. The second stage utilizes a Reinforcement Learning (RL)-based algorithm for the dynamic deployment of VNFs and their dedicated-active backups onto network nodes based on network conditions. A key innovation is the "deffer" method, where dedicated backups are not immediately deployed. Instead, their activation is decided based on the instantaneous network state. The RL agent learns the optimal timing for deploying or delaying backups, adapting to network dynamics.

Network topology significantly influences the reliability of SFCs. In \cite{r6}, the authors enhance network reliability by employing serial and parallel placement models for primary VNFs and their dedicated-active backups. Paper \cite{r6}, distinguishes between node failure probability (hardware) and the failure probability of the VNF deployed on it (software). When multiple VNFs (primary or backup) are placed on the same node, their placement is considered serial; placement on different nodes is parallel. In a serial arrangement of identical connected VNFs, hardware failure in any one leads to the failure of all. Thus, backups of a primary VNF should not be serially connected to it. The paper formulates an optimization problem to minimize deployment costs and maximize the minimum reliability of SFCs, considering edge resource constraints. 

In \cite{r7}, paper framework addresses backups by considering node structural correlation to avoid simultaneous failures in primary VNFs of SFC. It uses a "node dependency factor" to place backups on independent nodes and employs shared reservation for resource efficiency among VNFs of different SFCs. A weighted allocation algorithm optimizes backup resource selection for reliability and resource utilization.
Paper \cite{r8} improves SFC reliability by breaking SFCs into shorter sub-chains, thus reducing the number of required dedicated backups, as failure probability increases with the number of VNFs.

Allocating resources to primary VNFs  and their backups incurs both operational/maintenance costs and increased energy consumption. In \cite{r9}, an RL-based approach is presented to simultaneously optimize cost, energy consumption, and reliability. The placement problem is modeled as a graph matching problem, mapping the resource requirements of each SFC onto the network graph. Candidate nodes for dedicated-active allocation to VNFs and their backups are selected based on minimizing network link usage and energy consumption. This method, named Cand-RL, combines greedy candidate node selection with an RL agent for the final placement decisions of primary VNFs and their backups. Stochastic Petri Net models are used to accurately evaluate the reliability achieved by this resource allocation and SFC placement, simulating the failure and recovery of network nodes and VNFs.

Paper \cite{r10} investigates the cost-effective and reliable provisioning of SFCs in dynamic request environments with limited computational and memory resources, considering heterogeneous hardware and software reliability. To address this, the RuleDRL algorithm is proposed. This algorithm combines deep deterministic policy gradient for managing delayed rewards with a method for prioritizing backup allocation to the least reliable VNFs, employing dedicated-active backup to minimize unavailability. RuleDRL dynamically determines the number and placement of primary VNFs and their backups based on dynamic SFC requests, while respecting resource constraints.

To reduce network costs for SFC placement, \cite{r11} uses multi-agent RL to optimize VNF configuration, traffic routing, and VNF deployment. It employs deep neural networks for routing and deployment agents and heuristics for VNF configuration. A key innovation is "delay compensation" by allocating more processing resources, enabling diverse routing paths. Reliability is enhanced by deploying minimal-resource backups without significantly increasing operational costs. Similarly, \cite{r12} uses deep Q-networks to minimize SFC placement costs and maximize the number of accepted SFCs with guaranteed QoS, while meeting reliability requirements.
The work in \cite{r13} focuses on improving reliability and reducing costs in 5G networks. Considering end-to-end latency, it designs a reliable framework for SFC deployment resilient to VNF failures. It allocates a dedicated-active backup chain for each SFC, providing each VNF with a dedicated-active backup.

In time-sensitive applications like real-time IoT services, meeting deadlines and ensuring low latency are critical. Multi-Access Edge Computing (MEC) enables data processing closer to data sources and end-users, reducing latency, enhancing scalability, and improving QoS for applications like IoT and SFCs by leveraging edge resources and supporting various access methods. In this context, \cite{r14} combines MEC and fuzzy logic to optimize latency, routing costs, and load balancing in the placement of dynamically arriving SFCs.

Hierarchical allocation of heterogeneous computing resources with varying capacities and costs enhances the efficiency of SFC execution. Studies \cite{r15} and \cite{r16} demonstrate its effectiveness in optimizing resource consumption and ensuring SFC execution within maximum allowed latency. Specifically, \cite{r16} explores optimal placement strategies for VNFs to reduce operational costs, proposing the dynamic deactivation of idle servers as an energy-saving measure.

To the best of our knowledge, prior works on SFC placement in telecommunication networks commonly employ a dedicated-active backup strategy for enhanced reliability, where each VNF has one or more dedicated-active backup nodes. This study introduces four distinct backup  allocation strategies tailored to SFC request characteristics, with dedicated-active being the simplest. The other three  offer better resource efficiency and flexibility. Furthermore, existing studies often focus on specific aspects like reliability, latency, or costs. However, none simultaneously consider a comprehensive set of key features: reliability, service latency constraints, operational/maintenance costs, resource efficiency, and heterogeneous infrastructure limitations. This study models and solves a more realistic, comprehensive, and complex problem by jointly considering all these parameters, detailed in the next section. Table \ref{T1} compares thirteen selected related works based on their consideration of these criteria.

\section{SYSTEM MODEL AND PROBLEM FORMULATION}

\begin{table}
	\begin{center}
		\caption{List of Symbols}
		\begin{tabular}{|l l|}
			\hline
			{\textbf{Symbol}} & {\textbf{Description}} \\
			$M$ & Number of server (node) categories \\
			$C_i$ & The $i$-th category \\
			$M_i$ & Number of nodes in $C_i$\\
			$w_i$ & Clock frequency of each node in $C_i$ \\
			$p_{i,a}$ & Cost of each active node in $C_i$ \\
			$p_{i,s}$ & Cost of each standby node in $C_i$ \\
			$f_{i,a}$ & Failure rate of each active node in $C_i$ \\
			$f_{i,s}$ & Failure rate of each standby node in $C_i$ \\
			$b_{k,j}$ & Number of backup(s) for $V_{k,j}$ in dedicated strategies \\
			$\tilde{b}_{i,k}$ & Number of backup(s) in $C_i$ for $S_k$ in shared strategies\\
			$S$  & The set of all SFCs \\
			$S_k$ & The $k$-th SFC in $S$ \\
			$N$ & Number of all nodes \\
			$N_k$ & The chain length of the $S_k$ \\
			$V_k$ & List of VNFs in $S_k$ \\
			$V_{k,j}$ & The $j$-th VNF in $S_k$ \\
			$L_k$ & List of computational loads in $S_k$ \\
			$L_{k,j}$ & Computational load that is needed by $V_{k,j}$ 
			\\
			$R_k$ & Required reliability of $S_k$ \\
			$B_k$ & Backup allocation strategy for $S_k$
			\\
			$T_k$ & Latency deadline of $S_k$ \\
			$\tau_k$ & Latency  of $S_k$  \\
\hline
		\end{tabular}
		\label{T2}
	\end{center}
\end{table}	
In this section, we provide a description of the heterogeneous
fog computing environment and the SFC workload. Subsequently, we outline the optimization problem in a
well-defined mathematical framework, aiming to minimize the
average delay and cost while meeting the SFCs' deadline and reliability constraints.
\subsection{System Model}
\begin{figure*}
	\centerline{\includegraphics[width=0.9\linewidth]{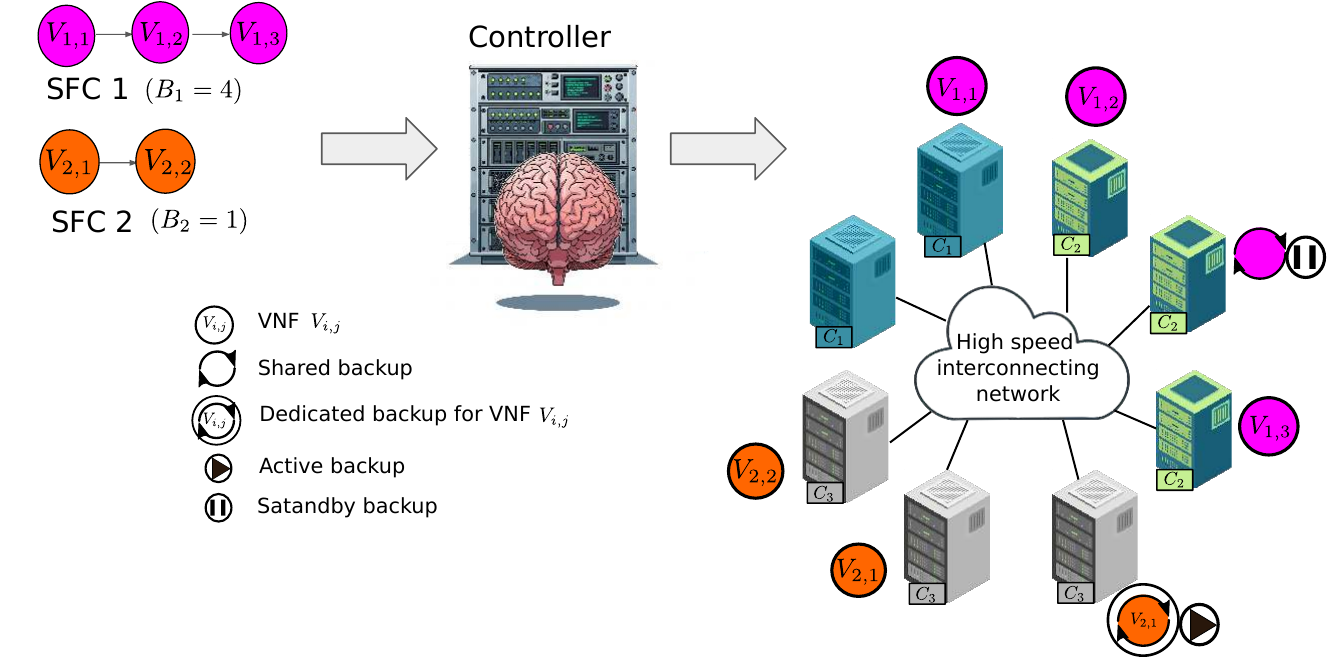}}
	\caption{A graphical example of the system model with two input SFCs. 
		}    
	\label{fig_sysmodel}
\end{figure*}
In this work, we assume a collection of $N$ interconnected servers (nodes), each capable of hosting a VNF \cite{r18}. The interconnecting network consists of high speed wired (e.g., switched Gbps Ethernet) or wireless (e.g., 5G) links. As a result, the transmission delay can be safely ignored.  Nodes are prone to failures due to hardware defects and/or software bugs \cite{r5}. Generally, the nodes are heterogeneous and can be classified into $ M $ distinct categories, $C_1,\cdots, C_M$ and $M \leq N$, according to their specifications.  The nodes in a category such as $C_i$, $i \in \{1, 2, ..., M\}$, is characterized by a tuple of ($M_i, w_i, p_{i,a}, p_{i,s}, f_{i,a}, f_{i,s}$). In this tuple, $M_i$ is the number of nodes in $C_i$ ($\sum_{i=1}^{M}M_i = N$), $w_i$ is the clock frequency, and $p_{i,m}$ , $m\in\{a,s\}$,  represents the cost of using a server based on its operation mode (active ($a$) or standby ($s$)). 
Moreover,  $f_{i,m}$ (where $m\in\{a,s\}$) is the failure rate, in either active ($ a $) or standby ($ s $) mode. The probability of failure, after $t$ units of operation time, follows an exponential distribution, with the cumulative distribution function (CDF) given by \cite{r19}:
\begin{flalign}
	F_{i,m}(t)=  1-\exp(-f_{i,m}t), m \in \{a, s\}
	\label{eq1}
\end{flalign}

On the other hand, the computational workload is modeled by a set of SFCs, denoted by $\mathcal{S} = {S_1, \dots, S_K}$, where each SFC $S_k$ has distinct VNF compositions and QoS requirements. Formally, $S_k$ is defined by the tuple $(N_k, V_k, L_k, T_k, R_k, B_k)$, where, $N_k$ is the chain length, $V_k = (V_{k,1}, \dots, V_{k,N_k})$ lists the VNFs in $S_k$, $L_k = (L_{k,1}, \dots, L_{k,N_k})$ specifies the CPU cycles (computational load) for each VNF $V_{k,j}$, $j=1,\dots,N_k$, $T_k$ is the latency deadline, $R_k$ is the reliability requirement, and $B_k$ is ``the backup strategy'' ensuring $R_k$ is met. Moreover, for simplicity, we assume the same service (holding) time for all SFCs. 

In this study, we examine four distinct backup strategies. Specifically, the backup strategy requested by SFC $S_k$ can be one of the following: \textit{dedicated-active} ($B_k=1$), \textit{dedicated-standby} ($B_k=2$), \textit{shared-active} ($B_k=3$), or \textit{shared-standby} ($B_k=4$). As detailed in the next section, these strategies differ in two key aspects: (1) whether backup nodes can be shared among multiple VNFs (dedicated or shared), and (2) the operational state of the backup nodes (active or standby).
For dedicated strategies ($B_k \in {1, 2}$), the number of dedicated backup nodes reserved for VNF $V_{k,j}$, allocated to a node of type $C_i$, is denoted by $b_{i,k,j}$. Conversely, for shared strategies ($B_k \in {3, 4}$), the number of shared backup nodes reserved for all VNFs $V_{k,j}$ allocated to $C_i$-type nodes is represented by $\tilde{b}_{i,k}$.
All notations  are summarized in Table \ref{T2}. 

Fig. \ref{fig_sysmodel} shows a graphical example of the system model with two input SFCs. SFC 1 (with three VNFs) requests a shared standby backup ($B_1 = 4$), while SFC 2 (two VNFs) requires a dedicated active backup strategy ($B_2 = 1$). The model incorporates three categories of fog servers ($C_1$, $C_2$, $C_3$). Here, it is assumed that the controller's placement decision deploys VNFs of SFC 1 across $C_1$ and $C_2$ servers, provisioning a shared standby backup for VNFs $V_{1,2}$ and $V_{1,3}$. In contrast, SFC 2 is deployed entirely on $C_1$ servers with a dedicated active backup for VNF $V_{2,1}$.

\subsection{Problem Definition}
Our objective is to minimize both execution delays and operational costs in SFC deployment by optimally allocating nodes to VNFs and their backups. This allocation must simultaneously satisfy reliability requirements and latency constraints. We therefore formulate the  {Delay, Cost, and Multi Reliability-aware} SFC placement problem over Heterogeneous Fog servers (\texttt{DCMR-HF}) as follows.

Let $\mathbf{x} =[x_{i,k,j}]$, $\mathbf{b} =[b_{i,k,j}]$, and $\tilde{\mathbf{b}} =[ \tilde b_{i,k}]$ denote the decision variables. Here, {$x_{i,k,j}\in\{0,1\}$} equals $1$ if VNF $V_{k,j}$ (including its backups) is assigned to a node of category $C_i$, and $0$ otherwise. The variable $b_{i,k,j}$ specifies the number of dedicated backup instances for $V_{k,j}$ in $C_i$, while $\tilde b_{i,k}$ denotes the number of shared backup instances assigned to  {SFC} $S_k$ in $C_i$. Moreover, $N_{i,k}=\sum_{j=1}^{N_k}x_{i,k,j}$ represents the number of VNFs assigned to category $C_i$. 
Accordingly the \texttt{DCMR-HF} is expressed as
\begin{align}
	\label{eq1_14}
	&\texttt{DCMR-HF}:\min_{\mathbf{x},\,\mathbf{b},\,\tilde{\mathbf{b}}} \quad 
	\alpha P_{normal} + \beta \tau_{normal} \\
	&\text{subject to}\notag\\ \quad 
	& \sum_{i=1}^{M} x_{i,k,j} = 1, 
	\quad \forall\, k \in \{1,\dots,K\},\; j \in \{1,\dots,N_k\} 
	\tag{a} \label{eq:a} \\
	& \sum_{k=1}^{K} \Bigg[ 
	N_{i,k}+\delta_{1,2}(B_k)\sum_{j=1}^{N_k}x_{i,k,j}b_{k,j}+ \label{eq:b} \tag{b} \\
	&
	\quad\delta_{3,4}(B_k)	\tilde{b}_{i,k}(1-\delta(N_{i,k})) 
	\Bigg] \leq M_i, 
	\quad \forall i \in \{1,\dots,M\} \notag\\
	& \Omega_k \geq R_k, 
	\quad \forall\, k \in \{1,\dots,K\} 
	\tag{c} \label{eq:c} \\
	& \tau_k \leq T_k, 
	\quad \forall\, k \in \{1,\dots,K\} 
	\tag{d} \label{eq:d} \\
	& x_{i,k,j} \in \{0, 1\}, b_{i,k,j}\in\mathbb{N}_0,\tilde{b}_{i,k}\in\mathbb{N}_0,\tag{e} \label{eq:e}\\
	&\quad  \forall\, i \in \{1,\dots,M\}, k \in \{1,\dots, K\}, j \in \{1,\dots,N_k\} 
	\notag
\end{align}
\noindent where 
 $\mathbb{N}0$ denotes the set of non-negative integers.
$\delta(z)$ and $\delta_{m,n}(z)$, are also indicator functions which return $1$ only if $z=0$ and $z\in\{m,n\}$, respectively, and $0$ otherwise. Moreover, $\Omega_k$ denotes the reliability of  $S_k$ (to be formally defined later), and
$\tau_k$ is  the latency (or execution) time of the SFC $S_k$ given by 
\begin{align}
	\tau_k =  \sum_{j=1}^{N_k} \sum_{i=1}^{M} \frac{L_{k,j}}{w_i} x_{i,k,j}.
\end{align}

Constraint (\ref{eq:a}) enforces one-to-one placement of each VNF, (\ref{eq:b}) ensures that the number of VNFs and their backups does not exceed the available nodes in each category, (\ref{eq:c}) enforces the reliability requirement of  each $S_k$, and (\ref{eq:d}) ensures that each $S_k$ finishes within its deadline.

In \texttt{DCMR-HF},  the optimization objective combines the normalized total delay and deployment cost, weighted by parameters $\alpha,\beta\in[0,1]$ with $\alpha+\beta=1$,  {$\tau_{normal}$ and $P_{normal}$} denote the normalized total delay and cost, defined as 
\begin{align}
	&\tau_{normal} = \frac{\sum_{k=1}^{K} \tau_k-\tau_{min}}{\tau_{max}}
	, \
	&P_{normal} = \frac{\sum_{k=1}^{K} P_k-P_{min}}{P_{max}}
\end{align}
where
$P_k$ is the deployment cost of $S_k$.
The values $\tau_{min}$ and $\tau_{max}$ denote, respectively, the minimum and maximum achievable total delay, obtained as	
\begin{align}
	&\tau_{min} = {\min_{\mathbf{x},\mathbf{b},\tilde{\mathbf{b}}}} \sum_{k=1}^{K}\tau_k, \
	&\tau_{max} = {\max_{\mathbf{x},\mathbf{b},\tilde{\mathbf{b}}}} \sum_{k=1}^{K}\tau_k,
\end{align}
both subject to constraints \eqref{eq1_14}(\ref{eq:a})–\eqref{eq1_14}(\ref{eq:e}).
Similarly, $P_{min}$ and $P_{max}$ denote the minimum and maximum achievable total cost:
\begin{align}
	&P_{min} = {\min_{\mathbf{x},\mathbf{b},\tilde{\mathbf{b}}}} \sum_{k=1}^{K}P_k, \
	&P_{max} = {\max_{\mathbf{x},\mathbf{b},\tilde{\mathbf{b}}}} \sum_{k=1}^{K}P_k,
\end{align}
again subject to \eqref{eq1_14}(\ref{eq:a})–\eqref{eq1_14}(\ref{eq:e}).   

In general, both $P_k$ and $\Omega_k$ depend on the backup strategy $B_k$ of SFC $S_k$. For the conventional case of dedicated-active backups ($B_k=1$), they take the form of
\begin{align}
	\label{eq_s1}
	\Omega_{k} = & \prod_{j=1}^{N_k}\sum_{i=1}^{M} \left(1 - [F_{i,a}(t)]^{b_{k,j}+1}\right)x_{i,k,j}
\end{align}
\begin{align}
	\label{eq_p1}
	P_{k}= & \sum_{j=1}^{N_k}\sum_{i=1}^{M} (b_{k,j}+1)x_{i,k,j}p_{i,a}. 
\end{align}
\noindent 
The specific formulations of $P_k$ and $\Omega_k$ for the remaining backup strategies are derived in the next section.   
\section {Reliability Strategies}
\label{S3}
This section presents various reliability strategies, along with the derivation of their corresponding reliability and cost formulas. These equations can then be incorporated into the objective function and constraint (\ref{eq:a}) of \texttt{DCMR-HF} problem.
\subsection{Strategy I: Dedicated-Active ($B_k=1$)}
A common approach to redundancy provisioning assigns dedicated active backups to each VNF, as described in \cite{r5}. In this setup, each VNF relies solely on its own set of backups, which are maintained in a ready-to-use state to minimize switchover time.
Based on this strategy, the reliability of SFC $S_k$, $\Omega_k$, is also  given by (\ref{eq_s1}).
It is assumed that all backup instances of a VNF are deployed on nodes within the same category as the node hosting the corresponding primary instance. The cost of serving $S_k$, $P_{k}$, is given by \eqref{eq_p1}.
\subsection{Strategy II: Dedicated-Standby ($B_k=2$)}
    Since backup nodes are idle until failures occur, they can remain in standby mode, which is suitable for SFCs with low  sensitivity to switching delays. In standby state, nodes have lower failure probabilities and incur reduced operational and maintenance costs, leading to both cost savings and improved energy efficiency.

This dedicated-standby strategy is more complex than dedicated-active. It requires modeling two types of failures: those of active nodes (hosting primary VNFs) and standby backups. When a standby node becomes active, its failure probability changes accordingly. These factors must be reflected in the reliability formulation. As shown in \cite{r20}, for systems with such behavior, component (here, VNF) reliability, with $b_{k,j}$ standby backups, can be calculated using {\eqref{eq4}}.
\begin{align}
	\Omega_{i,k,j} =& \frac{1}{b_{k,j}! f_{i,s}^{b_{k,j}}}\sum_{n=0}^{b_{k,j}} (-1)^n \binom{b_{k,j}}{n}\times\nonumber\\
	& \exp[-(f_{i,a} + n f_{i,s})t] \prod_{m=0, m \neq n}^{b_{k,j}} (f_{i,a} + m f_{i,s})&& \label{eq4}
\end{align}
Equation \eqref{eq4} calculates the reliability, or availability, of the VNF $V_{k,j}$ in the SFC $S_k$ using nodes of category $C_i$, based on the holding time $t$. In this equation, the parameter $b_{k,j}$ represents the number of backup nodes dedicated to the VNF $V_{k,j}$. Based on the component reliabilities in (\ref{eq4}), the reliability of $S_k$ takes the form of 
\begin{align}
	\Omega_{k}= & \prod_{j=1}^{N_k}\sum_{i=1}^{M}\Omega_{i,k,j}x_{i,k,j} & \label{eq5}
\end{align}
and the maximum placement cost of the  VNFs  and their corresponding backups is given by
\begin{align}
	P_{k}= & \sum_{j=1}^{N_k}\sum_{i=1}^{M}(p_{i,a}+b_{k,j}p_{i,s})x_{i,k,j}.  \label{eq6}
\end{align}
\subsection{Strategy III: Shared-Active ($B_k=3$)}
For VNFs deployed on servers within the same category, backups can be organized as a shared pool, allowing any backup to replace any failed primary VNF {\cite{r19}}. Assuming active backups, it is straightforward to show that the group reliability of all VNFs from SFC  $S_k$ running on category $C_i$ with $\tilde{b}_{i,k}$ shared backups is given by
\begin{align}
	\Omega_{i,k}^{(a)} = &\sum_{m=N_{i,k}}^{N_{i,k} + \tilde{b}_{i,k}} \binom{N_{i,k} + \tilde{b}_{i,k}}{m}\times\nonumber\\& (1-F_{i,a}(t))^m (F_{i,a}(t))^{N_{i,k} +\tilde{b}_{i,k} - m}.&& \label{eq7}
\end{align} 
The overall reliability and cost for the $S_k$ under this shared active backup strategy, are respectively given by 
\begin{align}
	\Omega_k= &\prod_{i=1, N_{i,k}\neq 0}^{M}\Omega_{i,k}^{(a)} & \label{eq8}
\end{align}
and
\begin{align}
	P_{k} = &\sum_{i=1,N_{i,k}\neq 0}^{M}(N_{i,k}+\tilde b_{i,k})p_{i,a}.& \label{eq9}
\end{align}

\subsection{Strategy IV: Shared-Standby ($B_k=4$)}
In Strategy IV, in contrast to Strategy III, the shared backup nodes are kept in standby mode, meaning they remain inactive until a failure occurs. 
To evaluate the group reliability of the VNFs belonging to service function chain $S_k$ that are deployed in node category $C_i$, we again adopt the reliability model 
proposed in \cite{r20}, to write the group reliability as:
\begin{align}
	& \Omega^{(s)}_{i,k} =  \frac{1}{\tilde b_{i,k}! f_{i,s}^{\tilde b_{i,k}}}\sum_{n=0}^{\tilde b_{i,k}} (-1)^n \binom{\tilde b_{i,k}}{n} \times\nonumber\\
	&\exp [-(N_{i,k}f_{i,a} + n f_{i,s})t] \prod_{m=0, m \neq n}^{\tilde b_{i,k}} (N_{i,k}f_{i,a} + m f_{i,s})  \label{eq10}
\end{align}
This expression calculates the probability that at least $ N_{i,k} $ nodes (among active and standby backups) remain operational during time $t$, considering the transition of backups from standby to active mode upon failure.
Accordingly, the overall reliability  for the $S_k$ which may span multiple node categories, is the product of the group reliabilities across all categories where $S_k$ has VNFs deployed and takes the form of

\begin{align}
	\Omega_k= &\prod_{i=1, N_{i,k}\neq 0}^{M}\Omega^{(s)}_{i,k}(t). & \label{eq11}
\end{align}
On the other hand, the corresponding cost is given by
\begin{align}
	P_{k} = &\sum_{i=1,N_{i,k}\neq 0}^{M}(N_{i,k}p_{i,a}+\tilde b_{i,k}p_{i,s}) & \label{eq12}
\end{align}

\section{Proposed Genetic Algorithms to Solve the \texttt{DCMR-HF} Problem}

In this section, we present two GA approaches for solving the \texttt{DCMR-HF} problem. These GAs are inspired by the principles of natural selection and evolution. A GA maintains a population of individuals (referred to as chromosomes), where each individual represents a potential solution to the problem. These chromosomes consist of genes, typically encoded as sequences of integers, and are evaluated using a fitness function that quantifies their effectiveness in solving the target problem.

The GA evolves the population over several generations. In each generation, a selection process identifies the fittest individuals, which then undergo genetic {operations, crossover and mutation,} to generate new offspring. Crossover combines segments of genetic material from two parent chromosomes, while mutation introduces random alterations to promote diversity. Through successive generations, the population ideally converges toward high-quality solutions by efficiently exploring the solution space.

\begin{figure}
	\centerline{\includegraphics[width=9cm]{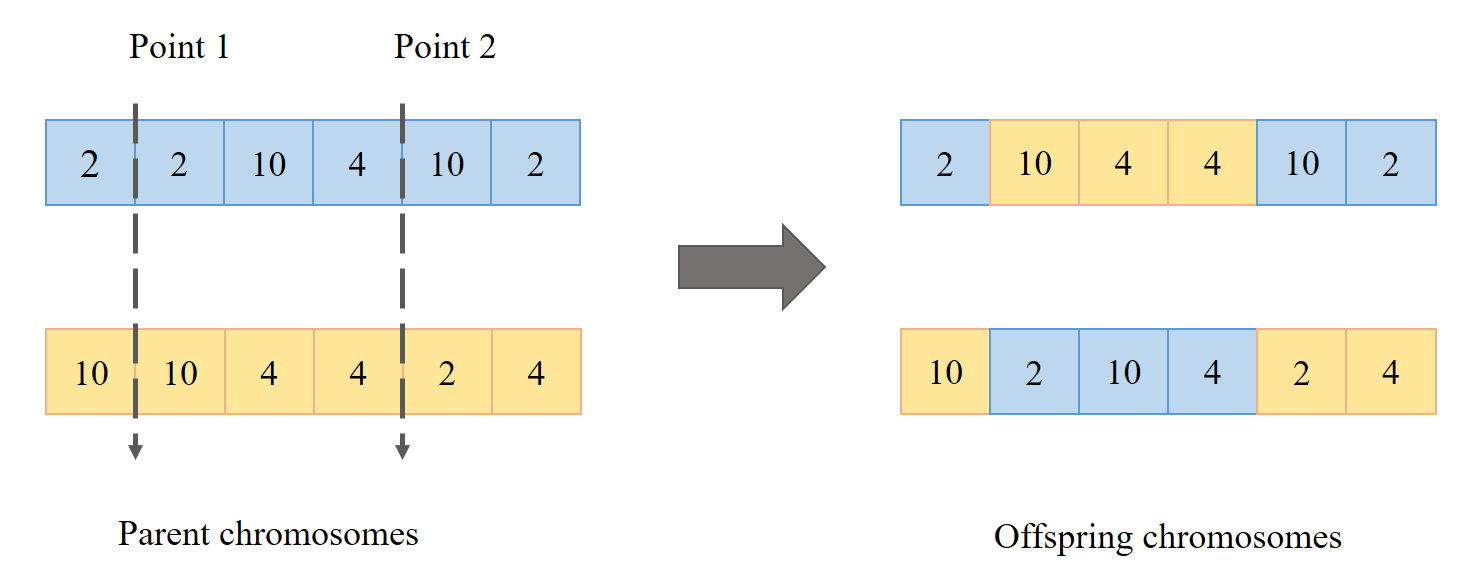}}
	\caption{Example of two points crossover.}
	\label{fig1}
\end{figure}

\begin{figure}
	\centerline{\includegraphics[width=9cm]{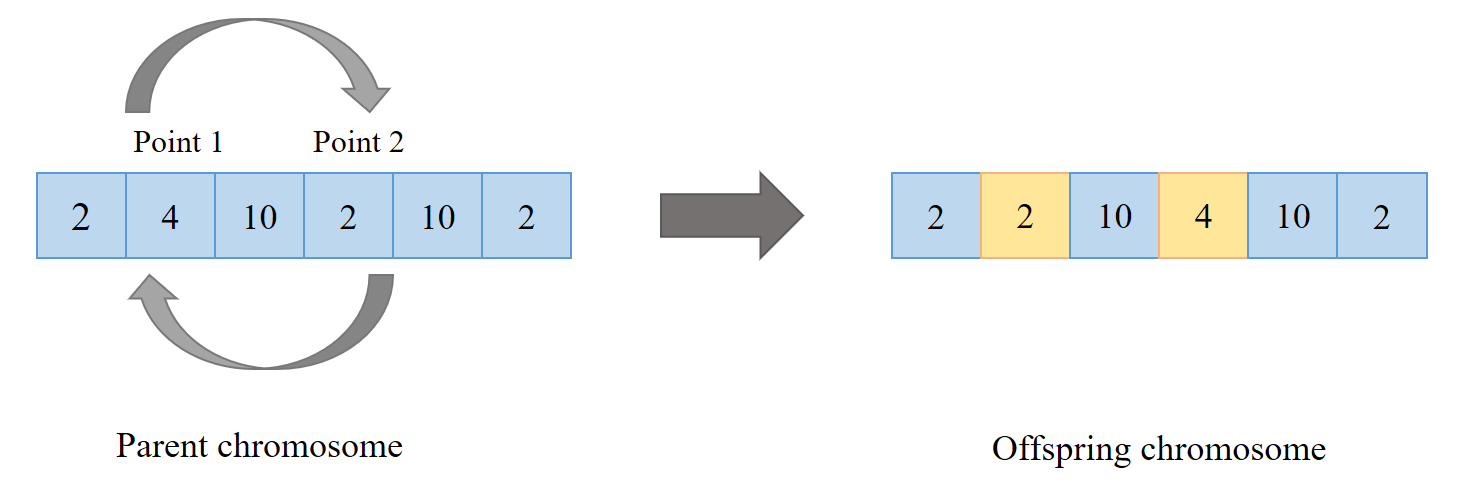}}
	\caption{Example of swap mutation.}
	\label{fig2}
\end{figure}

Various strategies can be employed for selection, crossover, and mutation. In this work, we utilize tournament selection \cite{r21}, two-point crossover, and swap mutation. Tournament selection involves running competitions among randomly chosen subsets of chromosomes, selecting the best performers to proceed. Two-point crossover randomly chooses two positions along the parent chromosomes and exchanges the {genes} between them, as illustrated in Fig. \ref{fig1}; this operation is equivalent to performing two single-point crossovers at distinct locations. Swap mutation, shown in Fig. \ref{fig2}, randomly selects two genes within a chromosome and swaps their values.

\subsection{Algorithm 1: GAP-GABA}
The first proposed algorithm, referred to as \textit{GA Placement and GA Backup Allocation} (GAP-GABA), represents each solution as a chromosome composed of  $N$ genes, where $N$ is the total number of nodes in the network. These genes are grouped according to the node categories: the first group corresponds to nodes in the first category, the second group to the second category, and so on. Each gene reflects the allocation status of the corresponding node in a potential solution. Each gene indicates the allocation status of its corresponding node in a candidate solution.

In this algorithm, first the VNFs of different SFCs are are \textit{globally} indexed from 1 to $\sum_k N_k$ where $N_k$ is the number of VNFs in SFC $k$. Genes then can take integer values in the range  $ 0 $ to $\sum_k N_k$.   A gene with a value of $0$ indicates that the node is inactive.  A non-zero value $i$ implies that the node either hosts the primary instance of  VNF with global index $i$ or serves as its backup. If multiple nodes in the same category have the same index  $i$, one is selected as the primary host, and the rest act as backups according to the backup strategy of the associated SFC. 

\begin{figure}
	\centerline{\includegraphics[width=5.8cm]{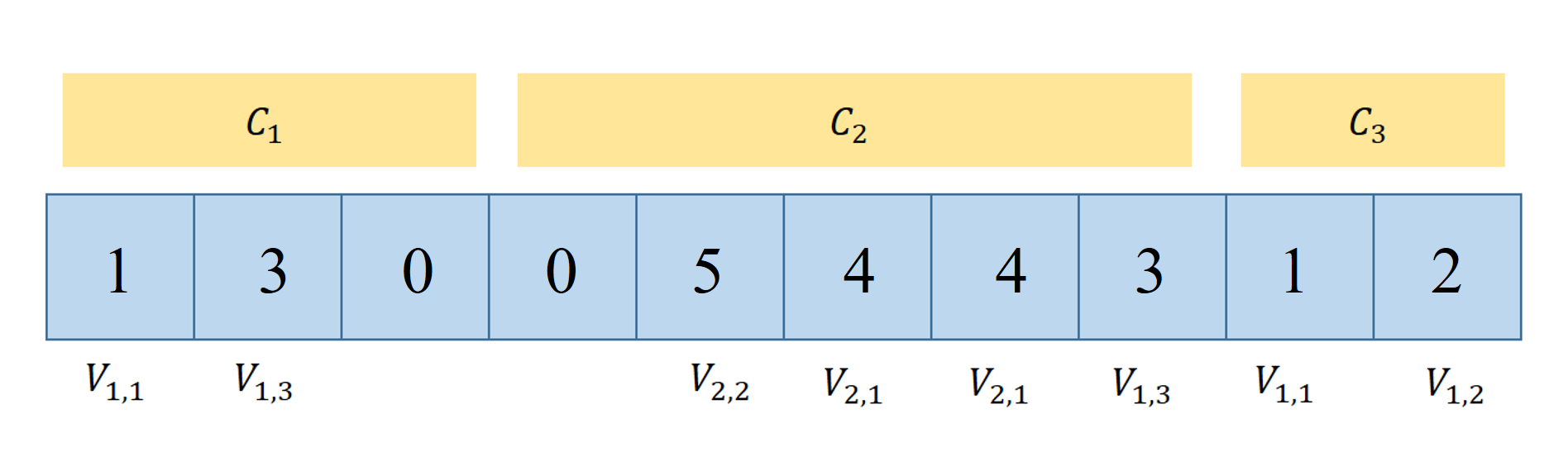}}
	\caption{An example of a GAP-GABA's chromosome.}
	\label{fig3}
\end{figure}

As an illustrative example, consider a network with ten nodes grouped into three categories: {$C_1$} with three nodes, $C_2$ with five, and $C_3$ with two. Two SFCs, $S_1$ and $S_2$ are to be deployed within this network. $S_1$ includes three VNFs, while  $S_2$ consists of two. In the GAP-GABA algorithm, all VNFs are first assigned global indices. For instance, the VNFs of $S_1$ (i.e., $V_{1,1}, V_{1,2}$, and $V_{1,3}$) are indexed as 1,2, and 3, and those of  $S_2$ (i.e., $V_{2,1}$ and $V_{2,2}$) as 4 and 5. Figure \ref{fig3} illustrates a sample chromosome that could be generated during the execution of the GAP-GABA algorithm, where the first three genes represent the nodes in category $C_1$, the next five correspond to the nodes in $C_2$, and the final two genes map to the nodes in $C_3$. 
This chromosome indicates that in category $C_1$,  one node is assigned to {$V_{1,1}$}, another to {$V_{1,3}$}, and the third remains inactive. In category $C2$, one node is unused, one is allocated to {$V_{2,2}$}, two nodes are assigned to {$V_{2,1}$}, and one to {$V_{1,3}$}. In category $C_3$, one node is assigned to {$V_{1,1}$} and the other to {$V_{1,2}$}. The assignment of two nodes in the same category $C_2$ to {$V_{2,1}$} is  valid; One node can serve as the primary host while the other acts as a backup. However, assigning {$V_{1,1}$} to nodes in two different categories $(C_1,C_3)$ violates the requirement that a VNF's primary and backup instances must be located within the same category. As a result, {$V_{1,1}$ must belong to a single category; therefore, either category $C_1$ or $C_3$ should be selected. The instance of $V_{1,1}$ in the non-selected category should be disregarded, and its corresponding node considered inactive.} 

 Procedure \ref{alg - gag} outlines the pseudo-code of the fitness function in the GAP-GABA algorithm.

\begin{algorithm}
	\caption{GAP-GABA fitness function}
	\label{alg - gag}
	\begin{algorithmic}[1]
		\renewcommand{\algorithmicrequire}{\textbf{Input}}
		\renewcommand{\algorithmicensure}{\textbf{Output}}
		\REQUIRE $ chromosome $
		\ENSURE  $ fitness $: fitness of $ chromosome $
		
			 \STATE\textit{\# Part 1: identifying SFC placement}
			\STATE $VNFCategory \leftarrow $  getVnfCategory$(chromosome)$
			\STATE $VNFRedundancy \leftarrow$  getVnfRedundancy$(chromosome$, $VNFCategory)$

			\STATE\textit{\#  Part 2: checking the allocations}

			\STATE $penalty \leftarrow 0$ 
			\FOR{$j$ from 1 to $\sum_k N_k$}
			\IF{$VNFCategory[j]$ == None}
			\STATE $penalty \leftarrow penalty + 1$
			\ENDIF
			\ENDFOR
			
			\STATE $totalExecutionTime \leftarrow 0$
			\STATE $totalCost \leftarrow 0$
			\FOR{$k$ from 1 to $K$}
			\STATE $SFC[k] \leftarrow$  getResource$(chromosome,k)$
			
			\STATE\textit{\# Part 3: checking reliability}
			\STATE $reliability \leftarrow$  getReliability$(SFC[k])$
			\IF{$reliability<R_k$}
			\STATE $penalty \leftarrow penalty + 1$
			\ENDIF
			
			\STATE\textit{\# Part 4: checking delay}
			\STATE $executionTime \leftarrow$ getExecutionTime$(SFC[k])$
			\STATE $totalExecutionTime \leftarrow totalExecutionTime + executionTime$
			\IF{$executionTime > T_k$}
			\STATE $penalty \leftarrow penalty + 1$
			\ENDIF
			
\textit{			\# Part 5: calculating costs}
			\STATE $cost \leftarrow $ getCost$(SFC[k])$ 
			\STATE $totalCost \leftarrow totalCost + cost$
			\ENDFOR
			
			\STATE $fitness \leftarrow \alpha * totalCost$ $ + \beta * totalExecutionTime$ 
			\hspace*{1.7cm}$+{\gamma} * penalty$
			
			\STATE return $fitness$

	\end{algorithmic}
	
\end{algorithm}

This fitness evaluation is structured into five key stages:

\begin{itemize}
	\item Part 1: Based on the gene values, the placement of each VNF and its corresponding backups within the node categories is determined. {If more than one VNF of the same type is deployed within a category, one instance is designated as the primary VNF, while the remaining instances are treated as backups. In the case of a dedicated backup strategies, these backups are associated with their respective VNFs. Conversely, under shared backup allocation strategies, the backups are associated with a group of VNFs belonging to the same SFC that includes that VNF type. In another scenario, if multiple VNFs of the same type are deployed across different categories, one category is selected randomly, and only the VNFs of that type within the selected category are retained. The VNFs of that type in the other categories are disregarded, and the nodes hosting them are considered inactive.}
	
	\item Part 2: {The algorithm verifies that all required VNFs are present in the chromosome. If any VNF is missing, a penalty is applied.}
	
	\item Part 3: For each SFC, reliability is calculated using the corresponding formula based on its specific backup allocation strategy. If the computed reliability does not meet the required threshold, an additional penalty is applied.
	
	\item Part 4: The execution delay of each SFC is computed, and if it exceeds the permitted threshold, an additional penalty is incurred.
	
	\item Part 5:  placement and backup costs are computed based on the SFC's backup allocation strategy.
\end{itemize}

Finally, the fitness score is calculated as a weighted sum of execution delay and allocation cost, with penalties added for violations such as missing VNFs, low reliability, or excessive delay. These penalties are scaled by a large constant {$\gamma$} ensuring that infeasible solutions receive higher (less favorable) scores.

\subsection{Algorithm 2: GAP-RABA}

\begin{figure}
	\centerline{\includegraphics[width=4cm]{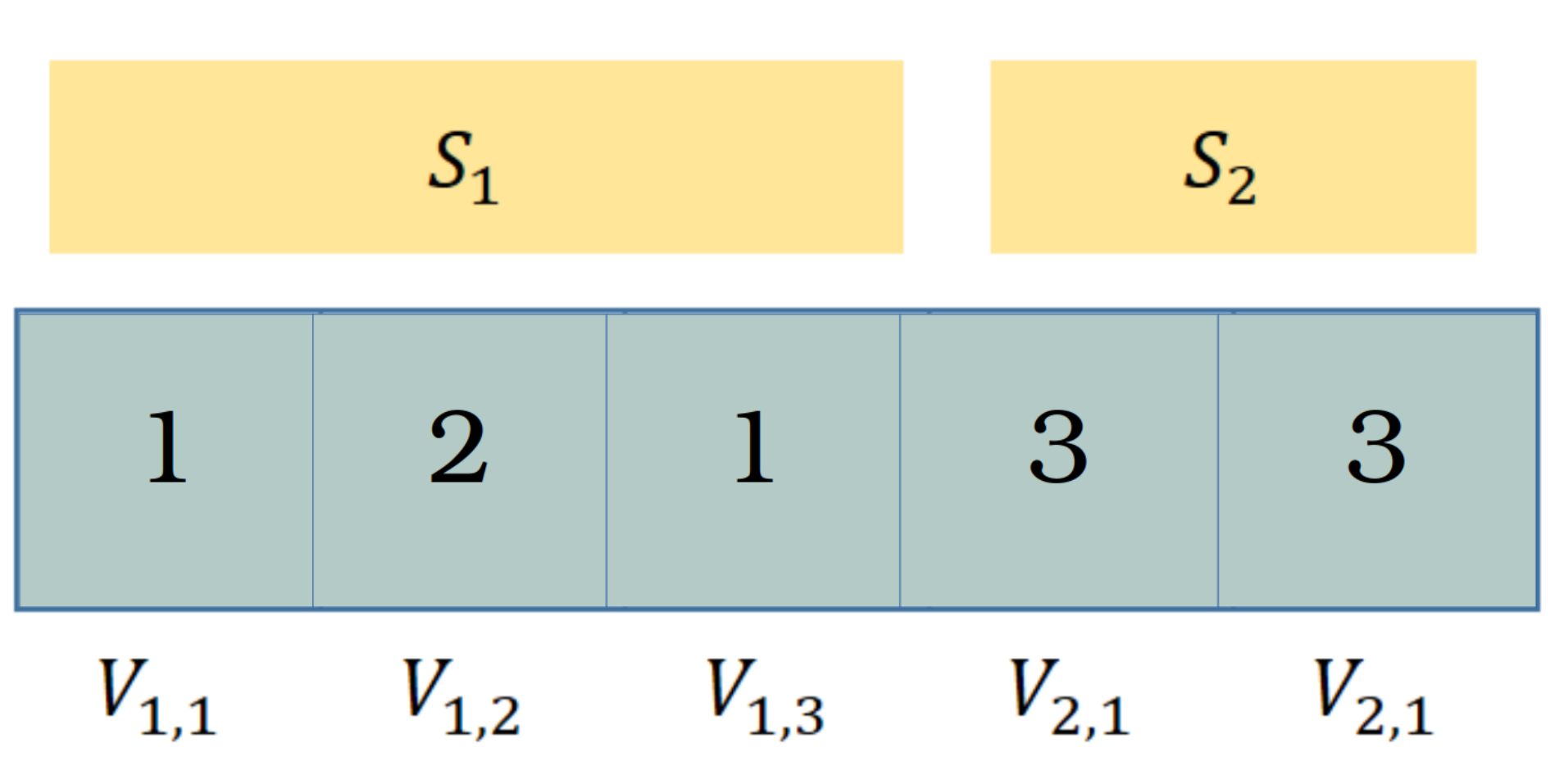}}
	\caption{Example of GAP-RABA's chromosome.}
	\label{fig4}
\end{figure}

The second algorithm proposed for solving the \texttt{DCMR-HF} problem, called \textit{GA Placement and Random Backup Allocation} (GAP-RABA), is also a GA-based metaheuristic, but it differs from GAP-GABA primarily in its chromosome design.
In GAP-RABA, each chromosome consists of $\sum_{k=1}^{K}N_k$ genes, where $N_k$ is the number of VNFs in SFC $S_k$. Each gene represents a VNF and its value indicates the node category where that VNF is placed. Genes are arranged consecutively per SFC: the first  $N_1$
genes correspond to the VNFs of $S_1$, the next $N_2$ ones to the $S_2$'s, and so on. 
For instance, in a network with three node categories and two SFCs $S_1$ with three VNFs and $S_2$ Figure \ref{fig4} shows a sample chromosome. Here, the first and the third VNFs of $S_1$ re placed in category $C_1$, the second in $C_2$, and both VNFs of $S_2$ in category $C_3$.

\begin{algorithm}
	\caption{GAP-RABA Random Backup Assignment and Fitness Function}
	\label{alg - gar}
	\begin{algorithmic}[1]
		\renewcommand{\algorithmicrequire}{\textbf{Input}}
		\renewcommand{\algorithmicensure}{\textbf{Output}}
		\REQUIRE $ chromosome $
		\ENSURE  $ fitness $: fitness of $ chromosome $

			\STATE\textit{\# Part 1: identifying SFC placement and assigning random priorities}
			\STATE  $SFC \leftarrow$ getDeployment$(chromosome)$
			\STATE $priority \leftarrow$ getPriority$(K)$
			\STATE $freeCapacity \leftarrow$ getFreeCapacity$(chromosome)$
			
			\STATE\textit{\# Part 2: backup allocation}
			
			\STATE $penalty \leftarrow 0$
			\STATE $backup \leftarrow [K]$
			\FOR{$k$ in $priority$}
			\STATE $backup[k],freeCapacity, reliability \leftarrow$ setBackup$(SFC[k],R_k, freeCapacity)$
			\IF{$reliability < R_k$}
			\STATE $penalty \leftarrow penalty + 1$
			\ENDIF
			\ENDFOR
			
			\STATE $totalExecutionTime \leftarrow 0$
			\STATE $totalCost \leftarrow 0$
			
			\FOR{$k$ form 1 to $K$}	
				\STATE 
			 \textit{\# Part 3: checking delay}
		 
			\STATE $executionTime \leftarrow$ getExecutionTime$(SFC[k])$
			\STATE $totalExecutionTime \leftarrow totalExecutionTime + executionTime$
			\IF{$executionTime>T_k$}
			\STATE $penalty \leftarrow penalty + 1$
			\ENDIF
\STATE			\textit{\textit{\# Part 4: calculating costs}}
			\STATE $cost \leftarrow $ getCost$(SFC[k])$ 
			\STATE $totalCost \leftarrow totalCost + cost$
			\ENDFOR
			
			\STATE $fitness \leftarrow \alpha * totalCost$ $ + \beta * totalExecutionTime$ 
			\hspace*{1.7cm}$+{\gamma} * penalty$
			
			\STATE return $fitness$
		
	\end{algorithmic}
	
\end{algorithm}

Procedure \ref{alg - gar} outlines the random backup assignment and fitness evaluation for each chromosome. It consists of four main steps, detailed as follows:
\begin{table*}[t]
	\centering
	\caption{{Dataset Parameters}}
	\renewcommand{\arraystretch}{1.1}
	\setlength{\tabcolsep}{4pt}
	\begin{subtable}[t]{0.9\textwidth}
		\caption{Node Parameters}
		\label{tab:sub1}
	\begin{tabular}{c c c c c c c c c}
		\hline
		\textbf{Node Parameters} & $M$ & $w_i$ & $p_a$ & $p_s$ & $f_a$ & $f_s$ & $N$ & $M_i$ \\
		\hline
		\textbf{Value(s)} & 3 & [5, 4, 1] & [25, 20, 5] & [2.5, 2, 0.5] & [0.008, 0.01, 0.04] & [0.0008, 0.001, 0.004] & 800 & [200, 300, 300] \\
		\hline
	\end{tabular}
	
	\end{subtable}
\vspace{0.5cm}\\
	\begin{subtable}[t]{0.9\textwidth}
		\caption{SFC Parameters}
		\label{tab:sub2}
	\begin{tabular}{c c c c c c c}
		\hline
		\textbf{SFC Parameters} & $K$ & $N_k$ & $L_k$ & $R_k$ & $T_k$ & $B_k$ \\
		\hline
		\textbf{Value(s)} & 10 & [5, 5, \ldots, 2, 5] & [[10, 20, \ldots, 9], \ldots, [20, 40, \ldots, 45]] & [0.99, 0.999, \ldots, 0.999] & [80, 10, \ldots, 100] & [1, 3, \ldots, 1] \\
		\hline
	\end{tabular}
\end{subtable}
	\label{T10}
\end{table*}

\begin{itemize}
	\item Part 1: The placement of all VNFs for all SFCs is extracted from the input chromosome. Since backup placement is also required and node availability within each category is limited, a priority order for backup allocation must be established. To this end, a random permutation of integers from 1 to $K$ is generated, defining the order in which SFCs will be processed during the backup assignment stage.
	
	\item Part 2: 
	Backups are assigned using the setBackup function, which employs a randomized approach. It begins by selecting a VNF placed in a category with available inactive nodes and assigns one of those nodes as a backup. The SFC’s reliability is then evaluated. If the required threshold is not met, another inactive node from a different category is randomly selected and assigned. This process repeats until either the reliability target is reached or no more inactive nodes are available. If reliability remains unsatisfied, a violation is recorded and penalized.
	\item Part 3: 
The execution time of each SFC is computed. If it exceeds the maximum allowable delay, a penalty is applied for the violation.
	
	\item Part 4: the cost imposed by each SFC is computed depending on the SFC's backup strategy. 
	
\end{itemize}

Finally, similar to Procedure \ref{alg - gag}, the fitness score is computed as a weighted sum of the total cost and execution time across all SFCs, with the penalty value added to the final result.

\section{Numerical Results}

In this section, we evaluate the performance of the proposed algorithms through simulation. All simulations were implemented in Python and executed on a machine equipped with an Intel Core i7 processor (2.9 GHz) and 8 GB of RAM. Portions of the {GA} were implemented using the PyGAD library \cite{r22}, which offers a flexible and feature-rich framework for developing evolutionary algorithms. 

We consider datasets with 2 to 3 categories of servers. The computation power of nodes also varies by category, with values between 1 and 5 units. Furthermore, the number of {nodes} in each category{,} failure rates in active mode, failure rates in standby mode, node cost in active mode, and node cost in the standby mode are chosen randomly in the range of  ($ 50-700 $), ($0.8\%-4\%$), ($0.08\%-0.4\%$), ($ 5-25 $), and ($ 0.5-2.5 $), respectively. Moreover, 
 there are {5 to 15} SFCs, each composed of 2 to 5 VNFs. The number of VNFs, their computational requirements, and the maximum acceptable delay per SFC are determined based on the specifications in \cite{r23}. The required reliability for each SFC ranges from $99\%$ to $99.9999\%$, supporting a wide spectrum of {applications from} moderately critical to ultra-reliable low-latency services 	\footnote{This broad range reflects the diversity of 5G service requirements, as emphasized in prior studies such as \cite{r24}, where services like industrial automation or autonomous driving demand extremely high reliability.} .  

The complete source code, along with {five} instances of datasets generated based on the above characteristics  are available on GitHub {\cite{r25}}. For the sake of brevity, in this paper, the results are reported just for one of the dataset instances. {Table \ref{T10}} summarizes the parameters of this dataset instance.   

The objective function uses scaling factors $\alpha = 0.65$ and $\beta = 0.35$. Furthermore, a one-year SFC retention period is assumed, aligning with the annual failure rates used for active nodes {\cite{r26}}. Table \ref{T9} presents the GA-specific parameters employed in the implementation of the GAP-GABA and GAP-RABA algorithms.
\begin{table} [t]
	\begin{center}
		\caption{GA parameters}
		\begin{tabular}{|c|c|}
			\hline
		Number of generations & $2000$ \\\hline
			Number of population & $400$ \\\hline
			Number of crossovers per generation & $380$ \\\hline
			Number of elites per generation & $100$ \\\hline
			Mutation rate & $10\%$ \\\hline
		\end{tabular}
		\label{T9}
	\end{center}
\end{table}

\subsection{Performance Evaluation Under Heterogeneous Backup Strategy Requests}
\begin{figure}
	\centerline{\includegraphics[width=\linewidth]{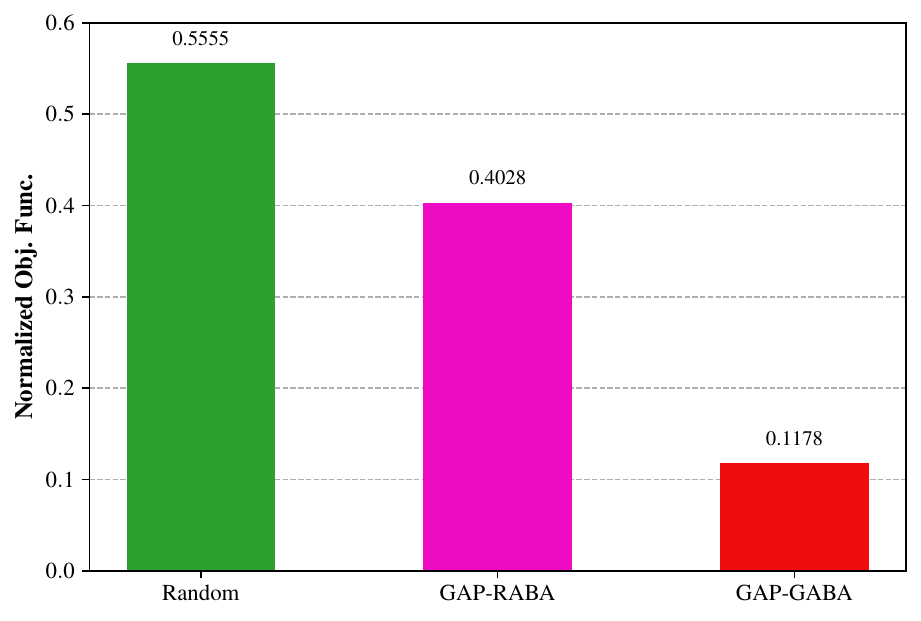}}
	\caption{Performance of different algorithms under random backup strategy scenario}
	\label{fig6}
\end{figure}

In this subsection we consider the case in which different SFCs may require heterogeneous backup strategies. Fig. \ref{fig6} illustrates the normalized objective function achieved by the evaluated algorithms under this condition. For comparison, we also include a baseline random-selection approach, which chooses any feasible solution uniformly at random and reports its resulting performance. As shown in Fig. \ref{fig6}, GAP-GABA consistently attains the lowest normalized objective value, indicating superior efficiency. Specifically, it yields approximately 71\% and 80\% reductions in the objective function compared to GAP-RABA and the Random algorithm, respectively, demonstrating a significantly more efficient performance. 

\begin{figure}
	\centerline{\includegraphics[width=\linewidth]{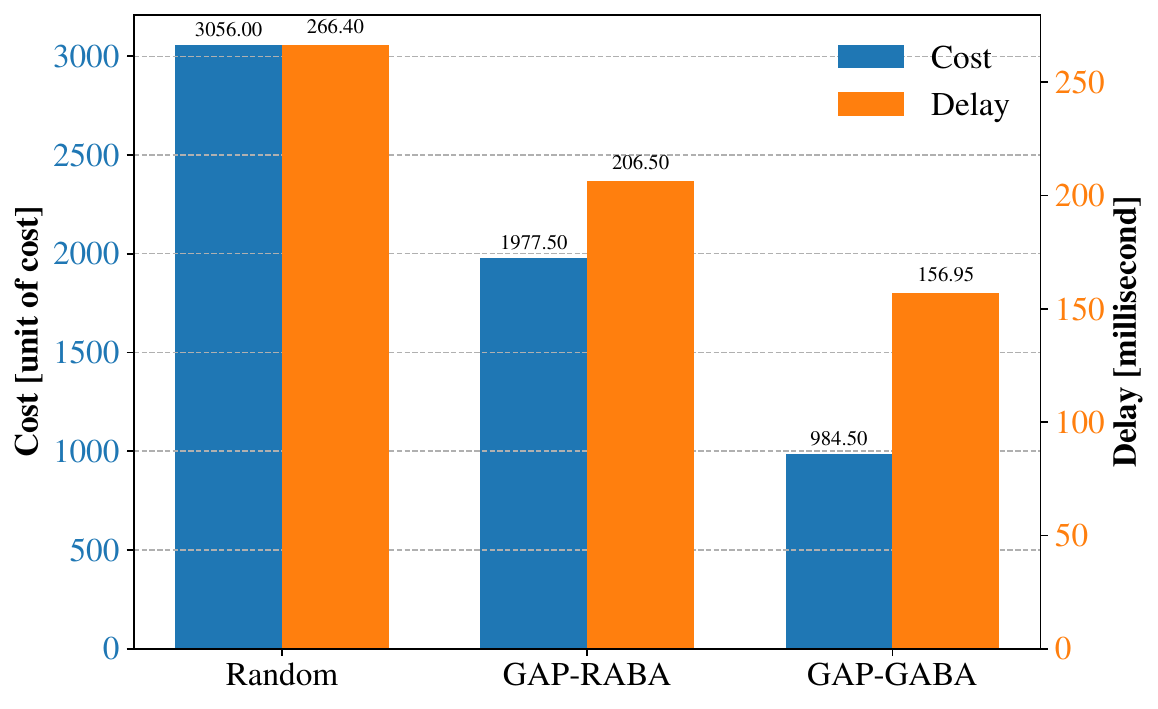}}
	\caption{The achieved cost and delay of different algorithms for the random backup strategy scenario}
	\label{fig_18_1}
\end{figure}

Fig. \ref{fig_18_1} further breaks down the normalized objective function into its constituent components, cost and delay, to provide a more detailed comparison. The results reveal that GAP-GABA not only achieves the best overall performance, but also consistently outperforms competing schemes across each individual metric. In particular, GAP-GABA reduces the cost component by approximately 50\% and 68\% relative to GAP-RABA and the Random method, respectively. Moreover, with respect to delay, it achieves improvements of nearly 24\% and 41\% compared to GAP-RABA and Random, respectively.
\subsection{Impact of Backup Strategy Selection on System Performance}
\begin{figure}
	\centerline{\includegraphics[width=\linewidth]{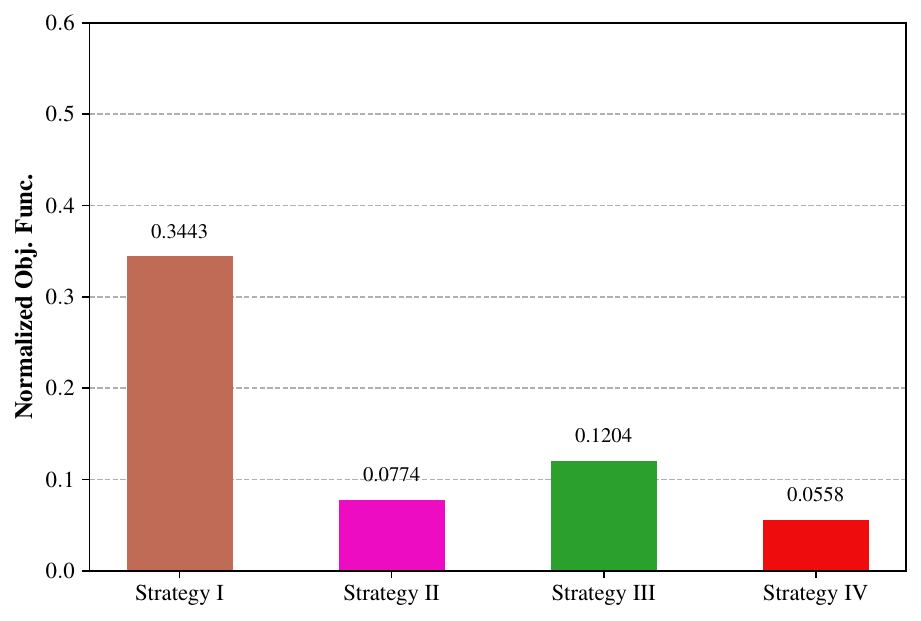}}
	\caption{Performance of GAP-GABA for different strategies}
	\label{fig2_27}
\end{figure}

\begin{figure}
	\centerline{\includegraphics[width=\linewidth]{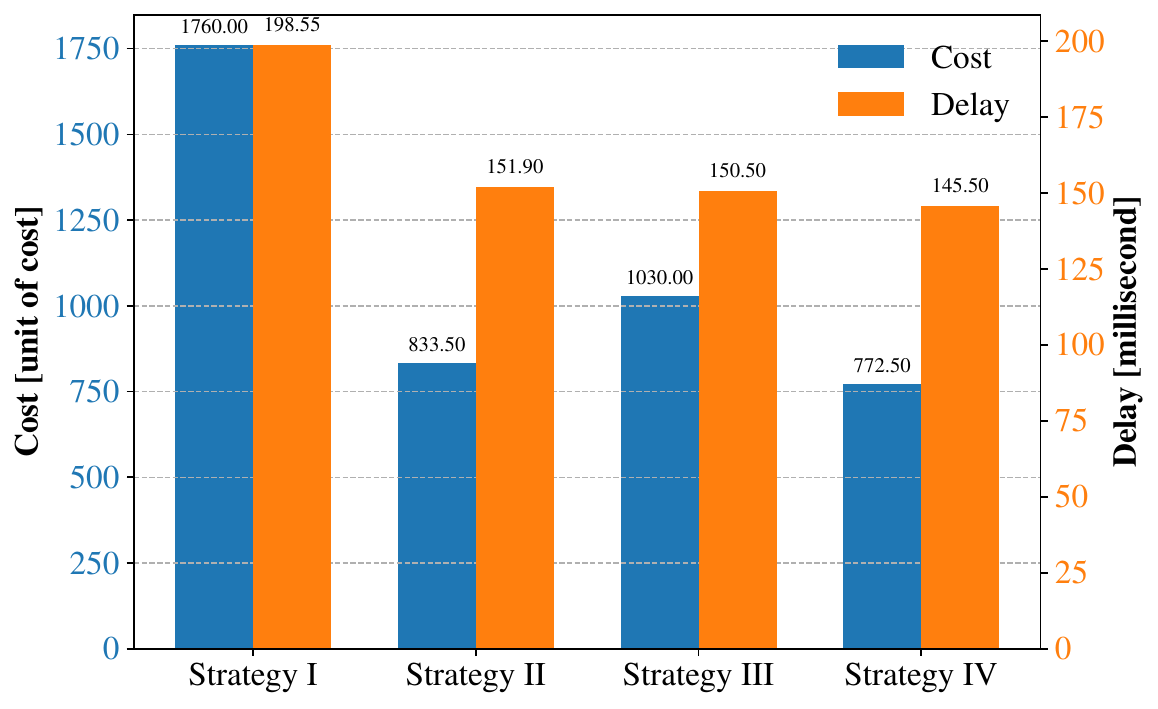}}
	\caption{The achieved cost and delay of GAP-GABA for different strategies}
	\label{fig_27_1}
\end{figure}

In this subsection, we investigate the impact of backup strategy selection on the overall performance of SFC placement. To isolate the effect of the strategy itself, we fix the placement algorithm to the proposed GAP-GABA method and evaluate four scenarios, each corresponding to a distinct backup strategy applied uniformly across all SFCs. The normalized objective function for these scenarios is illustrated in Fig.~\ref{fig2_27}. As shown, Strategy~IV yields the most favorable performance, achieving reductions of approximately 0.84\%, 28\%, and 54\% compared to Strategies~I, II, and III, respectively.  
To provide deeper insight into the performance gap, Fig.~\ref{fig_27_1} breaks down the normalized objective function into its cost and delay components. Strategy~IV demonstrates superiority in both metrics. Specifically, in terms of operational cost, Strategy~IV achieves reductions of approximately 56\%, 7\%, and 25\% compared to Strategies~I, II, and III, respectively. Likewise, with respect to delay, it improves performance by about 27\%, 4\%, and 3\% over the same strategies. These results confirm that Strategy~IV offers a more balanced trade-off between cost efficiency and latency reduction, leading to an overall enhancement in SFC placement performance.

\section{CONCLUSION}
In this paper, we revisited the problem of SFC placement with backup reservation to enhance resilience against hardware and software failures. Our key contribution was the introduction and evaluation of more efficient backup strategies beyond the conventional dedicated–active approach. Specifically, we examined shared–active, dedicated–standby, and shared–standby strategies, which, despite their foundation in reliability theory, have been largely overlooked in reliability-aware SFC placement.
We formulated the placement as an INLP to jointly minimize cost and delay, accommodating heterogeneous SFC backup requirements. Two GA-based solutions were proposed to address computational complexity. Results show that enabling both backup sharing and standby operation substantially improves performance, with the shared–standby strategy achieving up to 84\% reduction in a combined delay–cost metric relative to dedicated–active.

Future work includes extending the model to dynamic arrivals, variable SFC holding times, and cross-SFC backup sharing, as well as refining reliability modeling to decouple VNF and node failures for a more realistic system representation.

\bibliographystyle{IEEEtran}
\bibliography{references}

@Article{r1,
AUTHOR = {Atlam, Hany F. and Walters, Robert J. and Wills, Gary B.},
TITLE = {Fog Computing and the Internet of Things: A Review},
JOURNAL = {Big Data and Cognitive Computing},
VOLUME = {2},
YEAR = {2018},
NUMBER = {2},
ARTICLE-NUMBER = {10},
URL = {https://www.mdpi.com/2504-2289/2/2/10},
ISSN = {2504-2289},
DOI = {10.3390/bdcc2020010}
}

@article{r2,
	title={A survey on service function chaining},
	author={Bhamare, Deval and Jain, Raj and Samaka, Mohammed and Erbad, Aiman},
	journal={Journal of Network and Computer Applications},
	volume={75},
	pages={138--155},
	year={2016},
	publisher={Elsevier}
}

@article{r3,
author = {Hawilo, Hassan and Shami, Abdallah and Mirahmadi, Maysam and Asal, Rasool},
year = {2014},
month = {09},
pages = {},
title = {NFV: State of the Art, Challenges and Implementation in Next Generation Mobile Networks (vEPC)},
volume = {28},
journal = {IEEE Network},
doi = {10.1109/MNET.2014.6963800}
}

@INPROCEEDINGS{r5,
  author={Jia, Junzhong and Yang, Lei and Cao, Jiannong},
  booktitle={IEEE INFOCOM 2021 - IEEE Conference on Computer Communications}, 
  title={Reliability-aware Dynamic Service Chain Scheduling in 5G Networks based on Reinforcement Learning}, 
  year={2021},
  volume={},
  number={},
  pages={1-10},
  doi={10.1109/INFOCOM42981.2021.9488707}
}

@article{r6,
author = {Nguyen, Thi-Thuy-Lien and Pham, Tuan-Minh and Pham, Linh Manh},
title = {Efficient Redundancy Allocation for Reliable Service Function Chains in Edge Computing},
year = {2022},
issue_date = {Jan 2023},
publisher = {Plenum Press},
address = {USA},
volume = {31},
number = {1},
issn = {1064-7570},
url = {https://doi.org/10.1007/s10922-022-09708-x},
doi = {10.1007/s10922-022-09708-x},
journal = {J. Netw. Syst. Manage.},
month = dec,
numpages = {31}
}

@ARTICLE{r7,
  author={Woldeyohannes, Yordanos Tibebu and Tola, Besmir and Jiang, Yuming and Ramakrishnan, K. K.},
  journal={IEEE/ACM Transactions on Networking}, 
  title={CoShare: An Efficient Approach for Redundancy Allocation in NFV}, 
  year={2022},
  volume={30},
  number={3},
  pages={1014-1028},
  doi={10.1109/TNET.2021.3132279}}

@ARTICLE{r8,
  author={Thiruvasagam, Prabhu Kaliyammal and Kotagi, Vijeth J. and Murthy, C Siva Ram},
  journal={IEEE Transactions on Cloud Computing}, 
  title={A Reliability-Aware, Delay Guaranteed, and Resource Efficient Placement of Service Function Chains in Softwarized 5G Networks}, 
  year={2022},
  volume={10},
  number={3},
  pages={1515-1531},
  doi={10.1109/TCC.2020.3020269}}

@article{r9,
author = {Santos, Guto Leoni and Endo, Patricia Takako and Lynn, Theo and Sadok, Djamel and Kelner, Judith},
title = {A reinforcement learning-based approach for availability-aware service function chain placement in large-scale networks},
year = {2022},
issue_date = {Nov 2022},
publisher = {Elsevier Science Publishers B. V.},
address = {NLD},
volume = {136},
number = {C},
issn = {0167-739X},
url = {https://doi.org/10.1016/j.future.2022.05.021},
doi = {10.1016/j.future.2022.05.021},
journal = {Future Gener. Comput. Syst.},
month = nov,
pages = {93–109},
numpages = {17}
}

@ARTICLE{r10,
  author={Zeng, Yue and Qu, Zhihao and Guo, Song and Tang, Bin and Ye, Baoliu and Li, Jing and Zhang, Jie},
  journal={IEEE Transactions on Services Computing}, 
  title={RuleDRL: Reliability-Aware SFC Provisioning With Bounded Approximations in Dynamic Environments}, 
  year={2023},
  volume={16},
  number={5},
  pages={3651-3664},
  doi={10.1109/TSC.2023.3281759}}

@INPROCEEDINGS{r11,
  author={Li, Congzhou and Wu, Zhouxiang and Khanure, Divya and Jue, Jason P.},
  booktitle={2025 International Conference on Computing, Networking and Communications (ICNC)}, 
  title={A Multi-Agent Reinforcement Learning Scheme for SFC Placement in Edge Computing Networks}, 
  year={2025},
  volume={},
  number={},
  pages={446-451},
  doi={10.1109/ICNC64010.2025.10994017}}

@INPROCEEDINGS{r12,
  author={Khezri, Hamed Rahmani and Moghadam, Puria Azadi and Farshbafan, Mohammad Karimzadeh and Shah-Mansouri, Vahid and Kebriaei, Hamed and Niyato, Dusit},
  booktitle={2019 IEEE Global Communications Conference (GLOBECOM)}, 
  title={Deep Reinforcement Learning for Dynamic Reliability Aware NFV-Based Service Provisioning}, 
  year={2019},
  volume={},
  number={},
  pages={1-6},
  doi={10.1109/GLOBECOM38437.2019.9013214}}

@ARTICLE{r13,
  author={Thiruvasagam, Prabhu Kaliyammal and Chakraborty, Abhishek and Mathew, Abin and Murthy, C. Siva Ram},
  journal={IEEE Transactions on Network and Service Management}, 
  title={Reliable Placement of Service Function Chains and Virtual Monitoring Functions With Minimal Cost in Softwarized 5G Networks}, 
  year={2021},
  volume={18},
  number={2},
  pages={1491-1507},
  doi={10.1109/TNSM.2021.3056917}}

@Article{r14,
AUTHOR = {Guo, Shuang and Du, Yarong and Liu, Liang},
TITLE = {A Meta Reinforcement Learning Approach for SFC Placement in Dynamic IoT-MEC Networks},
JOURNAL = {Applied Sciences},
VOLUME = {13},
YEAR = {2023},
NUMBER = {17},
ARTICLE-NUMBER = {9960},
URL = {https://www.mdpi.com/2076-3417/13/17/9960},
ISSN = {2076-3417},
DOI = {10.3390/app13179960}
}

@article{r15,
author = {Tashtarian, Farzad and Zhani, Mohamed Faten and Fatemipour, Bita and Yazdani, Delaram},
title = {CoDeC: A Cost-Effective and Delay-Aware SFC Deployment},
year = {2020},
issue_date = {June 2020},
publisher = {IEEE Press},
volume = {17},
number = {2},
issn = {1932-4537},
url = {https://doi.org/10.1109/TNSM.2019.2949753},
doi = {10.1109/TNSM.2019.2949753},
journal = {IEEE Trans. on Netw. and Serv. Manag.},
month = jun,
pages = {793–806},
numpages = {14}
}

@article{r16,
author = {Nguyen, Minh and Dolati, Mahdi and Ghaderi, Majid},
title = {Deadline-Aware SFC Orchestration Under Demand Uncertainty},
year = {2020},
issue_date = {Dec. 2020},
publisher = {IEEE Press},
volume = {17},
number = {4},
issn = {1932-4537},
url = {https://doi.org/10.1109/TNSM.2020.3029749},
doi = {10.1109/TNSM.2020.3029749},
journal = {IEEE Trans. on Netw. and Serv. Manag.},
month = dec,
pages = {2275–2290},
numpages = {16}
}

@Article{r17,
AUTHOR = {Qu, Hua and Wang, Ke and Zhao, Jihong},
TITLE = {Reliable Service Function Chain Deployment Method Based on Deep Reinforcement Learning},
JOURNAL = {Sensors},
VOLUME = {21},
YEAR = {2021},
NUMBER = {8},
ARTICLE-NUMBER = {2733},
URL = {https://www.mdpi.com/1424-8220/21/8/2733},
PubMedID = {33924460},
ISSN = {1424-8220},
DOI = {10.3390/s21082733}
}

@Article{r18,
AUTHOR = {Xu, Yansen and Kafle, Ved P.},
TITLE = {An Availability-Enhanced Service Function Chain Placement Scheme in Network Function Virtualization},
JOURNAL = {Journal of Sensor and Actuator Networks},
VOLUME = {8},
YEAR = {2019},
NUMBER = {2},
ARTICLE-NUMBER = {34},
URL = {https://www.mdpi.com/2224-2708/8/2/34},
ISSN = {2224-2708},
DOI = {10.3390/jsan8020034}
}

@book{r19,
  author    = {Igor A. Ushakov},
  title     = {Optimal Resource Allocation: With Practical Statistical Applications and Theory},
  publisher = {Wiley},
  year      = {2013},
  url       = {https://books.google.com/books?id=a88AqAuYIaoC},
  isbn      = {9781118593037}
}

@ARTICLE{r20,
  author={She, J. and Pecht, M.G.},
  journal={IEEE Transactions on Reliability}, 
  title={Reliability of a k-out-of-n warm-standby system}, 
  year={1992},
  volume={41},
  number={1},
  pages={72-75},
  doi={10.1109/24.126674}}

@inproceedings{r21,
  author    = {Ying Fang and Jing Li},
  title     = {A Review of Tournament Selection in Genetic Programming},
  booktitle = {Advances in Computation and Intelligence},
  editor    = {Zhenhua Cai and Changyin Hu and Zengqiang Kang and Yang Liu},
  series    = {Lecture Notes in Computer Science},
  volume    = {6382},
  pages     = {181--192},
  publisher = {Springer},
  address   = {Berlin, Heidelberg},
  year      = {2010},
  doi       = {10.1007/978-3-642-16530-6_21}
}

@article{r22,
  title={Pygad: An intuitive genetic algorithm python library},
  author={Gad, Ahmed Fawzy},
  journal={Multimedia Tools and Applications},
  pages={1--14},
  year={2023},
  publisher={Springer}
}

@INPROCEEDINGS{r23,
  author={Onsu, Murat Arda and Lohan, Poonam and Kantarci, Burak and Janulewicz, Emil and Slobodrian, Sergio},
  booktitle={GLOBECOM 2024 - 2024 IEEE Global Communications Conference}, 
  title={A New Realistic Platform for Benchmarking and Performance Evaluation of DRL-Driven and Reconfigurable SFC Provisioning Solutions}, 
  year={2024},
  volume={},
  number={},
  pages={85-91},
  doi={10.1109/GLOBECOM52923.2024.10901609}}

@inproceedings{r24,
	title={Ultra-reliable low-latency in 5G: A close reality or a distant goal?},
	author={Maghsoudnia, Arman and Vlad, Eduard and Gong, Aoyu and Dumitriu, Dan Mihai and Hassanieh, Haitham},
	booktitle={Proceedings of the 23rd ACM Workshop on Hot Topics in Networks},
	pages={111--120},
	year={2024}
}

@misc{r25,
  author       = {Negin Doostar},
  title        = {Beyond Dedicated Active Redundancy Strategies for Reliable SFC Placement in Fog Computing},
  year         = {2025},
  howpublished = {\url{https://github.com/NeginDoostar/Beyond-Dedicated-Active-Redundancy-Strategies-for-Reliable-SFC-Placement-in-Fog-Computing}},
  note         = {Accessed: Oct. 30, 2025},
}

@misc{r26,
  author       = {{Statista}},
  title        = {Annual failure rates of servers},
  year         = {2024},
  howpublished = {\url{https://www.statista.com/statistics/430769/annual-failure-rates-of-servers/}},
  note         = {Accessed: Oct. 31, 2025},
}

\end{document}